\pdfoutput=1

\documentclass[fleqn,10pt]{wlscirep}
\usepackage[utf8]{inputenc}
\usepackage[T1]{fontenc}
\usepackage{graphicx,url}

\title{Top influencers can be identified universally by combining classical centralities}

\author[1,*]{Doina Bucur}
\affil[1]{University of Twente, Department of Computer Science, Drienerlolaan 5, 7522 NB Enschede, The Netherlands}
\affil[*]{d.bucur@utwente.nl}

\begin{abstract}
Information flow, opinion, and epidemics spread over structured networks. When using individual node centrality indicators to predict which nodes will be among the top influencers or spreaders in a large network, no single centrality has consistently good ranking power. We show that statistical classifiers using two or more centralities as input are instead consistently predictive over many diverse, static real-world topologies. Certain pairs of centralities cooperate particularly well in statistically drawing the  boundary between the top spreaders and the rest: local centralities measuring the size of a node's neighbourhood benefit from the addition of a global centrality such as the eigenvector centrality, closeness, or the core number. This is, intuitively, because a local centrality may rank highly some nodes which are located in dense, but peripheral regions of the network---a situation in which an additional global centrality indicator can help by prioritising nodes located more centrally. The nodes selected as superspreaders will usually jointly maximise the values of both centralities. As a result of the interplay between centrality indicators, training classifiers with seven classical indicators leads to a nearly maximum average precision function (0.995) across the networks in this study. 
\end{abstract}
\begin{document}

\flushbottom
\maketitle

\thispagestyle{empty}

% ______________________________________________________________________________________________________________
\section*{Introduction}

Social influence, news, as well as infectious disease diffuse in society, following links drawn between participants by frequent contact, mutual interests, collaboration, communication, or transportation. The influence of a single node in such a network measures the extent to which the node, acting as the seed of a multi-hop diffusion process, will activate the rest of the network (this is the cascade size in the domain of online social networks, and the attack rate or outbreak size in epidemiology). Even assuming that the network links are known and the process of diffusion can be modelled or measured, \emph{predicting the top influential nodes} when knowing their nodes' topological \emph{centrality indicators} remains difficult, also because of the diversity and size of social contact topologies. This study shows, on real-world social networks, that in many networks the joint values of two or more (dissimilar) node centrality indicators are enlightening for the influence of the node, and that good combinations are between one local centrality which measures the size of the node's neighbourhood and one global centrality: a variant of the eigenvector centrality, closeness, or the node's core number. We illustrate with examples how the addition of such a second centrality to the prediction process is beneficial on some networks, and show that simple, interpretable statistical models can be machine-learnt in a supervised fashion on two or more centrality indicators, with almost universally good results across many real networks and network categories. 

Most prior studies predict top influencers by a \emph{ranking method}\cite{mariani2020network}: the nodes in a network are ranked according to a single centrality, with the top assumed to be the best influencers. No single centrality is consistent in performance across realistic case studies. The degree centrality was found a weak predictor in early studies, over both simulated and measured diffusion\cite{watts2007influentials,cha2010measuring}. With the susceptible–infectious–recovered (SIR)\cite{anderson1979population} diffusion model and also with measured diffusion, the top $f=5\%$  spreaders in a small number of networks were better predicted by their core numbers than by degree or betweenness centrality\cite{kitsak2010identification,pei2014searching}. The predictive power of the core number was later shown to not generalise, for SIR influence at or above the epidemic threshold. In road networks, the core number correlated little with the spreading ability of a node, while in social networks the degree and core number were either equally predictive\cite{de2014role}, or variably predictive with $f$\cite{liu2015core}. Over a test suite of ten networks, the eigenvector centrality was on average better than the core number\cite{macdonald2012spreaders}. Refinements of classical centrality indicators were also defined\cite{lu2011leaders,garas2012k,chen2013identifying,zeng2013ranking,liu2013ranking,chen2014path,de2014role,liu2015improving,liu2016identify,radicchi2016leveraging,wang2017ranking,li2018identification}. Ideas around combining centrality indicators into a predictor of influence started in 2011. A metric equal to the betweenness centrality of a node, divided by a power of its degree\cite{comin2011identifying} was used to recognise the seed of a diffusion process, but was not successful on a real-world topology. By 2020 (the time of this writing), some methods\cite{mo2015evidential,liu2015node,bian2017identifying,rodrigues2019machine,zhao2020machine} were not applied beyond relatively small or few networks, and also provide no explanation or intuition for the results. A scalable method based on graph neural networks\cite{fan2020finding} was black-box and cannot explain its decision. More interpretable approaches\cite{madotto2016super,ibnoulouafi2018m} aggregated the individual rankings or values of two or more centralities, with coefficients based on the correlations between the rankings, or the information entropy of a centrality. This obtained recognition rates above 0.7 in 16 networks, with 9–18\% improvement over the best single ranking in five of these networks, and a lower 1–5\% in the rest\cite{madotto2016super}; still, no explanation was given for the strength of centrality combinations. 

\begin{figure}[htb]

    \includegraphics[width=\textwidth]{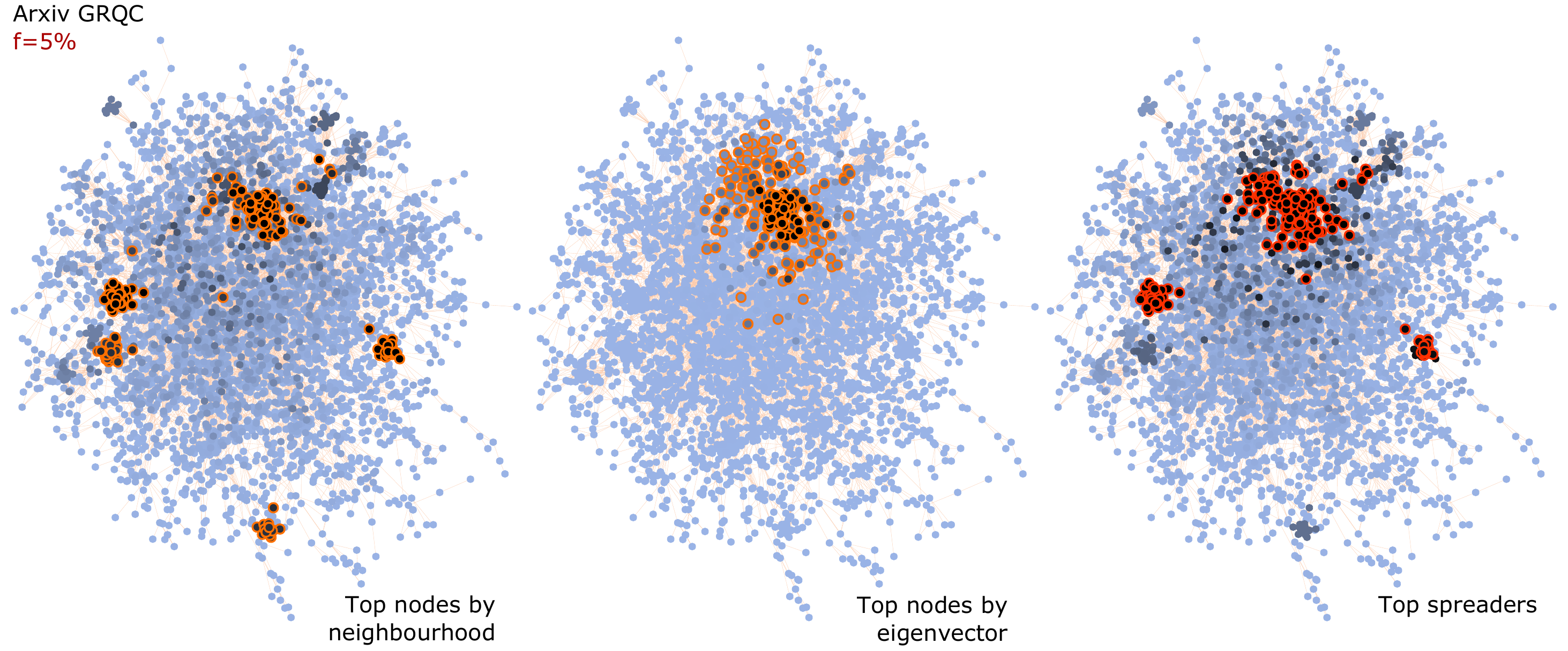}
    
    \caption{Comparing the location of the top nodes as ranked by (left) neighbourhood size, (centre) eigenvector centrality, and (right) SIR spread size at the epidemic threshold for the coauthorship network Arxiv GRQC. The network layout is force-directed. The colour of the nodes in each panel shows the value of that metric: darker nodes have higher centrality values or spread size. The top $f=5\%$ of the nodes in each case are encircled.}
    \label{fig:example-Arxiv_GRQC}
\end{figure}

From 2020, two studies gained insights which converge on the same basic idea. Over all non-isomorphic small networks (up to 10 nodes), one normalized spectral centrality (PageRank or Katz centrality) together with degree (or another measure of network density) predicted spread sizes for single nodes well, compared with exact computations of SIR spread sizes\cite{bucur2020beyond}. For the related problem of maximising collective influence, PageRank plus metrics related to the node's degree and neighbourhood brought 2-5\% improvement compared to the baseline greedy heuristic\cite{erkol2019systematic}. Here, we aim for more general answers: are there other good combinations of classical centralities? can one explain the added value of a centrality? does the predictive power of a combination of centralities generalise across many topologies? We give an early example in Fig.~\ref{fig:example-Arxiv_GRQC}, for the 4,158-node coauthorship network Arxiv GRQC. When predicting the top $f=5\%$ influencers by the size of its neighbourhood (the sum of the degrees of the nearest neighbours, on the left), the resulting nodes (encircled) form clusters distributed in the network. When doing the same by the eigenvector centrality (centre), the top nodes are instead local to one cluster. Neither of these solutions entirely coincides with the correct set of top nodes by their spread size, but reasoning with both sets of data leads to a good prediction. The true top spreaders by the SIR diffusion model at the epidemic threshold are shown on the right: these are located in and around only that subset of the clusters with a large neighbourhood which have \emph{also} marginally higher eigencentrality values due to being in or close to the high-eigencentrality cluster. (Fig.~\ref{fig:two-centrality-examples-Eigenvector} will provide more detail.)

We study a large and diverse set of real-world test networks of sizes between 1,000 and 70,000 nodes, assuming complete knowledge of the links in the network. The predictive power of two or more centrality indicators is measured by training a supervised statistical classifier on sample nodes from each network. The ground truth for the influence of any node is estimated accurately via the simulation of the SIR diffusion model with that node as the seed of diffusion---possible here since there is one seed, unlike in studies on collective influence, where an approximate greedy heuristic must instead be used as a baseline\cite{erkol2019systematic}. The target of the classification is then a binary variable which shows whether the node is in the true top $f\%$ of spreaders. While the results are diverse across the set of networks, we find six universally good pairs between one local centrality which measures the density of the node's extended neighbourhood and one global centrality (eigencentrality or PageRank, closeness, core number), and give an intuition for why they complement each other well. With all seven classical centralities, the average precision function is close to perfect (0.995) and the average recognition rate is 0.921.

The practical use of these results is two-fold. The method of supervised classification can be ported to any new network where the assumption of complete knowledge about the links is satisfied. For a more realistic estimation of node influence, empirical diffusion data\cite{cha2010measuring,pei2014searching}, when available, can replace the mathematical model of diffusion. More importantly, the basic principles of centrality pairing can help with the design of more effective centrality indicators or ranking algorithms, and can improve the understanding of diffusion outcomes in social networks. 

% ______________________________________________________________________________________________________________
\section*{Results}

% Summarise study design: networks, diffusion model
We run an empirical study over 60 real-world examples of static network topologies (listed in Table~\ref{tab:networks} in Methods). The networks are directed, unweighted, and fall into six categories: human social networks (separately, online or offline), human networks formed by professional coauthorship or online communication, computer networks, and physical infrastructure. The influence of a node is the SIR spread size when the node is the seed of diffusion, estimated via Monte Carlo simulation (see Methods). Analyses are shown in this section for the SIR influence at the epidemic threshold $\lambda_c$ for every network; they hold also above the epidemic threshold, at $1.5\cdot\lambda_c$ (with numerical results for these shown in the Supplementary Information). 

We study seven classical centrality indicators and their combinations, as follows. 
\begin{itemize}
    \item Local metrics, simple to compute, reflect the density of a node's neighbourhood: the \emph{degree}, \emph{neighbourhood} (the sum of the degrees of direct neighbours), and \emph{two-hop neighbourhood} (the sum of the degrees of neighbours exactly two hops away).
    \item The \emph{core number} results from k-shell decomposition.
    \item Distance-based centralities, such as \emph{closeness} and \emph{betweenness}, reflect the importance of nodes by their link distances in the network. Of these two popular centralities, in prior studies on the SIR model, betweenness showed weak predictiveness both as a ranker of nodes in large networks\cite{kitsak2010identification,de2014role} and also in combinations with other centralities on small networks\cite{bucur2020beyond}. We thus study here the closeness centrality.
    \item Normalised spectral centralities: \emph{PageRank} and \emph{eigenvector} centrality.
\end{itemize}

% ______________________________________________________________________________________________________________
\subsection*{The predictive power of single centralities is inconsistent across networks}

\begin{figure}[htb]

    \includegraphics[width=\textwidth]{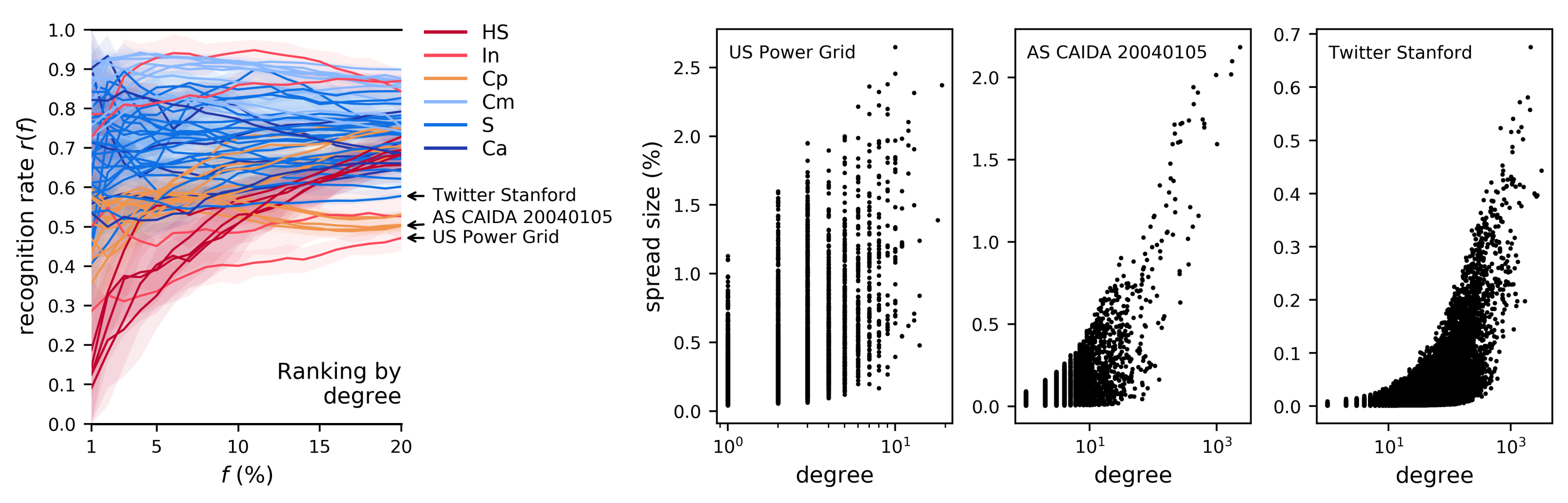}
    
    \caption{(left) The recognition rate by \textbf{degree}, across $f$, for all networks, at the SIR epidemic threshold. Each data line corresponds to a network, with the 95\% confidence interval shown as a shaded, partly transparent area. The network categories are \textit{Ca} (Coauthorship, 6 networks), \textit{Cm} (Communication between people, 11 networks), \textit{Cp} (Computer, 11 networks), \textit{HS} (offline Human Social, 5 networks), \textit{In} (Infrastructure, 4 networks), \textit{S} (online Social, 23 networks). (right) Degree-influence scatterplots for three of the worst-case networks.}
    \label{fig:recognition-rate-degree}
\end{figure}

We first show that the ability of any one centrality indicator to predict the top spreaders across a large number of network cases is too variable to be of universal practical use. Take a network of $N$ nodes, $f$ a fraction, and the task of selecting the best $fN$ spreaders in the network. The standard ranking method has each centrality rank the nodes in this network; the top $fN$ nodes by this ranking are put forward as best spreaders\cite{kitsak2010identification,pei2014searching,de2014role,liu2015core,macdonald2012spreaders} (see Methods). The predictive power of the degree centrality is shown in Fig.~\ref{fig:recognition-rate-degree}, across all networks, at the epidemic threshold. This is measured via the \emph{recognition rate} (also called recall) $r(f)$: the fraction of correctly identified top spreaders (Eq.~\ref{eq:pm1} in Methods); the 95\% confidence interval around $r(f)$ is shown as a shaded area. In Fig.~\ref{fig:recognition-rate-degree}, for each of the three categories of networks with lowest recognition rates at $f=20\%$, the worst-case network is named. The degree-influence scatterplots, also in Fig.~\ref{fig:recognition-rate-degree}, show the reason: a correlation between degree and influence does exist even in these worst cases, but with too wide a variance of influence per degree for accurate ranking.

\begin{figure}[htb]

    \includegraphics[width=\textwidth]{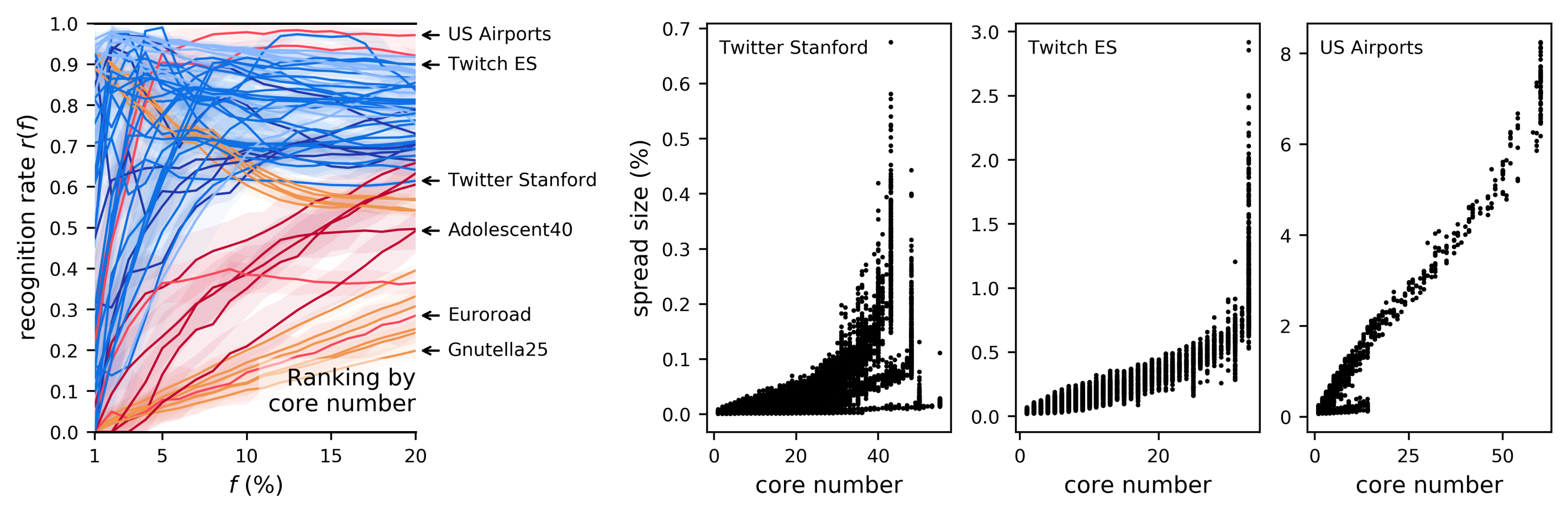}
    
    \caption{As Fig.~\ref{fig:recognition-rate-degree}, but with the \textbf{core number} as the ranker.}
    \label{fig:recognition-rate-corenumber}
\end{figure}

Compared to the degree, the performance of the core number as a ranker is much less consistent across networks (Fig.~\ref{fig:recognition-rate-corenumber}). The same cause holds for the three worst-case networks marked in the figure: all have few k-shells (between 1 and 5), so the core number by itself it not a discriminative variable for a ranking task. In the very worst case (as in the case of Gnutella25), the network has a single k-shell, so predicting the top spreaders by ranking the nodes in the network is the same as doing a random draw. In Fig.~\ref{fig:recognition-rate-corenumber}, three more networks are marked, for which ranking by core number gives good recognition rates at $f=20\%$, but poor rates when $f<5\%$. The scatterplots between core number and influence show the cause. The nodes with the highest core number in the Twitter Stanford network are poor spreaders; a topological reason for this was found in a prior study focused on the core number\cite{liu2015core}: the most effective core in the network depends not only on its core number, but also on its connectivity to other cores. Even in other topologies, in which high core numbers do correlate with wide spreading (as is the case for Twitch ES and US Airports), the highest core contains many nodes of very variable influence, so the core number alone is not a sufficiently discriminative variable when $f$ is low.

Neither the degree centrality nor the core number are universally better than the other across the network space. If the core number can be a more accurate ranker in some cases (Fig.~\ref{fig:recognition-rate-corenumber} shows values of $r(f)$ closer to 1 for the core number, as was also found in prior studies on selected topologies\cite{kitsak2010identification,pei2014searching}), it is also a poor predictor in absolute terms when $f<5\%$ for many networks, and also across all $f$ values when the network doesn't have a strong core structure. For online human networks (categories \textit{Ca}, \textit{Cm}, and \textit{S} in this study), and with $f>5\%$, Figs.~\ref{fig:recognition-rate-degree}--\ref{fig:recognition-rate-corenumber} show the two centralities to be comparable, with the core number marginally better; in general, as recognised before\cite{macdonald2012spreaders,de2014role,liu2015core}, the predictive power of the core number is not consistently better than the degree centrality for SIR influence.

\begin{figure}[htb]

    \includegraphics[width=\textwidth]{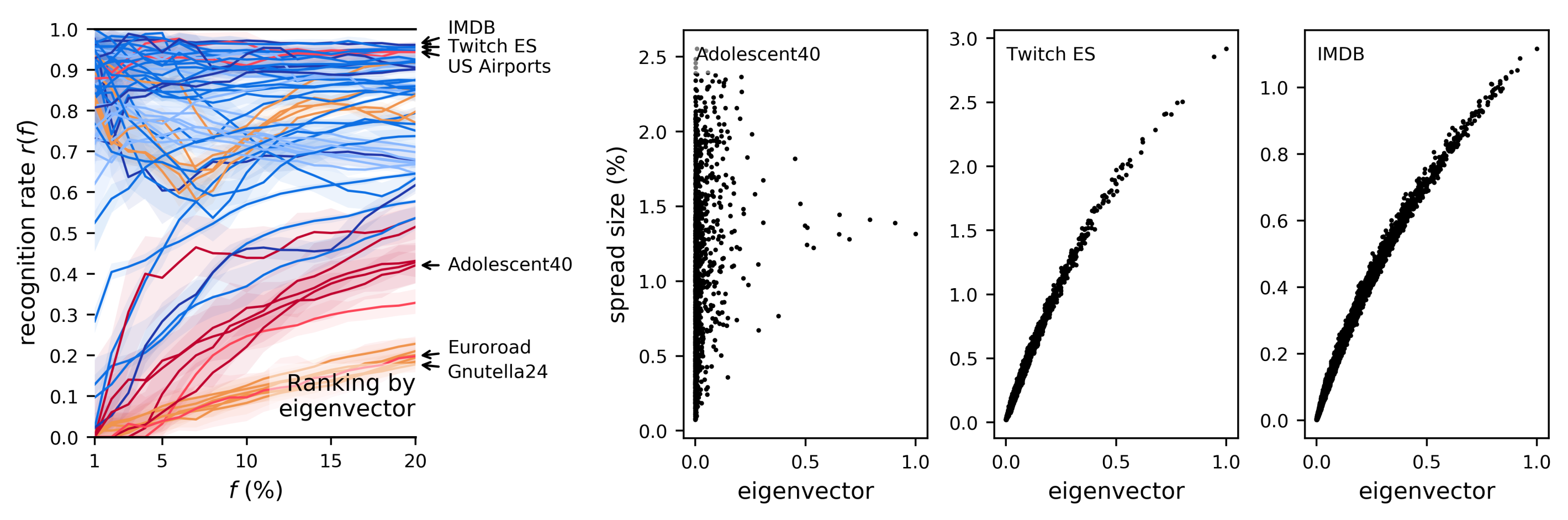}
    
    \caption{As Fig.~\ref{fig:recognition-rate-degree}, but with the \textbf{eigenvector centrality} as the ranker.}
    \label{fig:recognition-rate-eigenvector}
\end{figure}

Another popular ranker, the eigenvector centrality was previously found (on average across a set of networks) more predictive than the core number\cite{macdonald2012spreaders}. By the summary in Fig.~\ref{fig:recognition-rate-eigenvector}, this is the case for low values of $f$, but there is still a wide variance between networks. In some cases (such as Gnutella24 and Euroroad, marked in the figure), the distribution of centrality values is such that ranking is not better than a random draw; in others, such as Adolescent40, there is little correlation between the centrality and influence, so the ranking remains poor. In the best of cases (for two of which scatterplots are shown in the figure), this correlation is strong, which explains why the eigenvector centrality can be a very good predictor across the range of $f$.

\begin{figure}[htb]
    \centering

    \includegraphics[width=.8\textwidth]{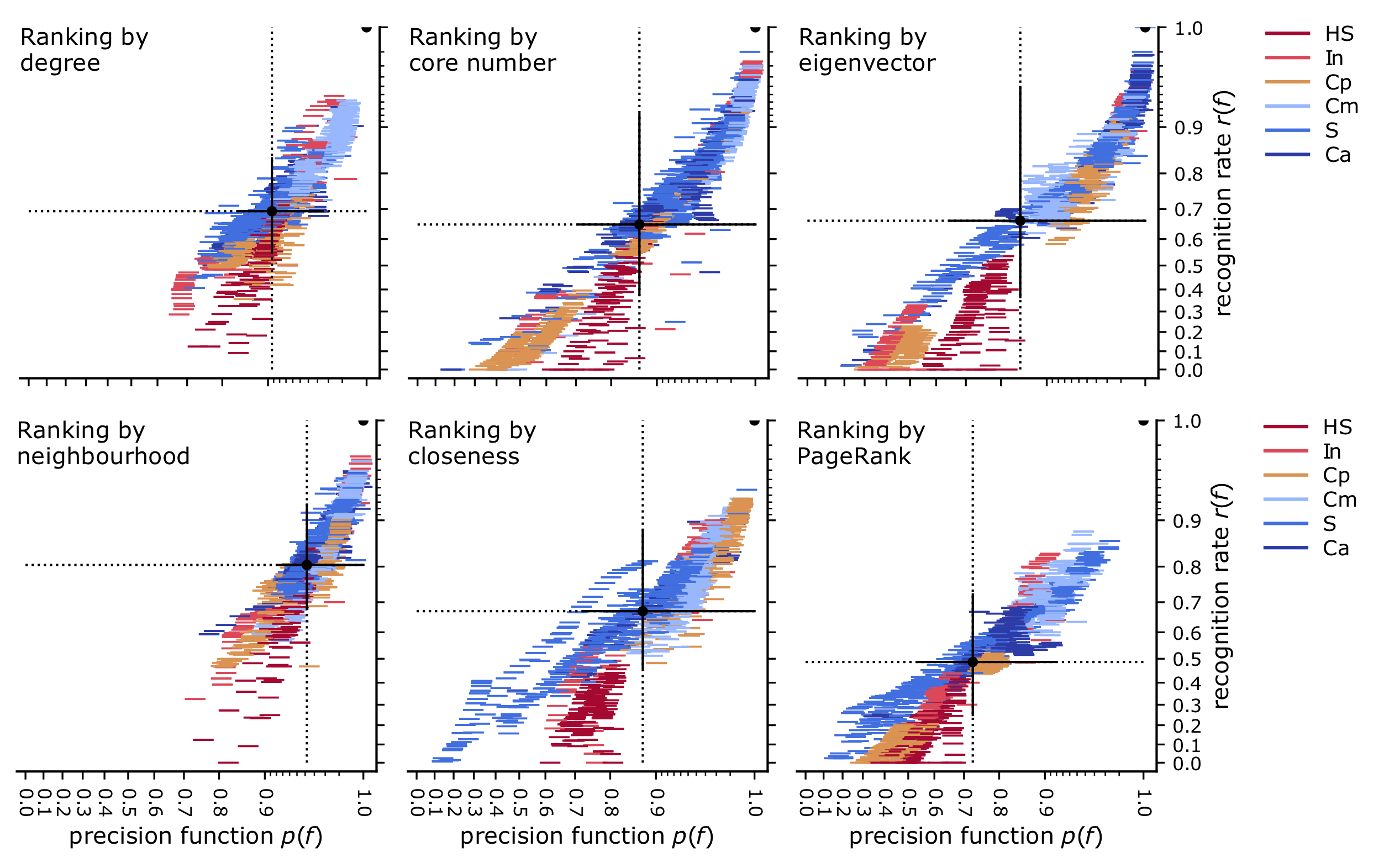}

    \caption{The success of \textbf{single-centrality ranking} at predicting spreaders, across all networks and values of $f$, at the SIR epidemic threshold. The scales are quadratic. Each data point (a horizontal line of fixed size) denotes a prediction task, and the colour shows the category of the network (listed in Table~\ref{tab:networks} in Methods). The centroid of the point cluster and the standard deviation on both axes are marked with a solid dot and lines. The point of perfect scores (1,1) is also marked with a half circle. The neighbourhood centrality is the best overall single ranker, with an average precision function of 0.962 and an average recognition rate of 0.804.}
    \label{fig:twometric-comparison-ranking}
\end{figure}

A second performance metric is also of interest: the \emph{precision function} $p(f)$ (Eq.~\ref{eq:pm1} in Methods), which compares the SIR influence of the predicted nodes with the SIR influence of the correct top spreaders. A $p(f)$ value close to 1 for a prediction task means that, regardless whether or not the exact top spreaders were identified, the influence of the nodes which were identified is close to that of the set of top spreaders---so $p(f)$ does not penalise node substitutions, if the substitutes are similar in terms of influence. For ranking by single centralities, the results for both the recognition rate and the precision function are shown in Fig.~\ref{fig:twometric-comparison-ranking}. Each data point marks the performance of a ranking task, over a given network, for a value of $f$ in $1, 2, \ldots 20\%$. (To make the data points visible despite many partial overlaps, each data point is a horizontal line; this line does not denote the uncertainty of the data, but is of fixed size.) The centroid of each data cloud summarises the performance of that centrality over this set of networks. Overall, the neighbourhood centrality makes for the best single ranker, with an average recognition rate of 0.804 and an average precision function of 0.962. The two-hop neighbourhood (not shown in the figure) is only slightly worse (on average 0.781 and 0.942, respectively). PageRank is the least accurate, with an average recognition rate of 0.487, and an average precision function of 0.727. This latter result is not entirely surprising: although widely used for ranking nodes in network structures\cite{lu2016vital}, PageRank was found before to not be a competitive predictor for measured diffusion in various networks\cite{macdonald2012spreaders,pei2014searching}.

Next, we show that certain pairs of centrality indicators have, together, sufficient topological information about network nodes to improve the accuracy of the prediction tasks. 

% ______________________________________________________________________________________________________________
\subsection*{Pairs of centralities combine into better predictors}

A statistical classifier is now trained with multi-variate data from part of the nodes in each network. The result is one trained classifier per network and fraction $f$. For training, a centrality is one input feature. The target variable (or class) is binary, and it shows whether or not a node is in the top fraction $f$ in the network by spread size. The two performance metrics for the classifiers are the same as for ranking tasks, with the difference that the recall $r(f)$ is now improved as the F1 score, which is the harmonic mean between the precision of classification and the recall (for motivation, see Methods, Eq.~\ref{eq:pm2}). 

Parsimonious statistical models are beneficial to gain clear intuition about the results. We report here the most \emph{interpretable} statistical models which have good performance: support-vector machine (SVM) with second-degree polynomials as kernels (see Methods), whose \emph{decision boundaries} between classes are simple to understand. We verified that other, higher-variance statistical models based on decision trees have similar performance (with numerical results for Random Forests shown in the Supplementary Information). We start with training SVM classifiers with two centralities, and show that, for certain network examples, certain pairs of centralities build on each other's strengths and obtain predictive models that are significantly better than either centrality alone. 

\begin{figure}[htb]

    \includegraphics[width=\textwidth]{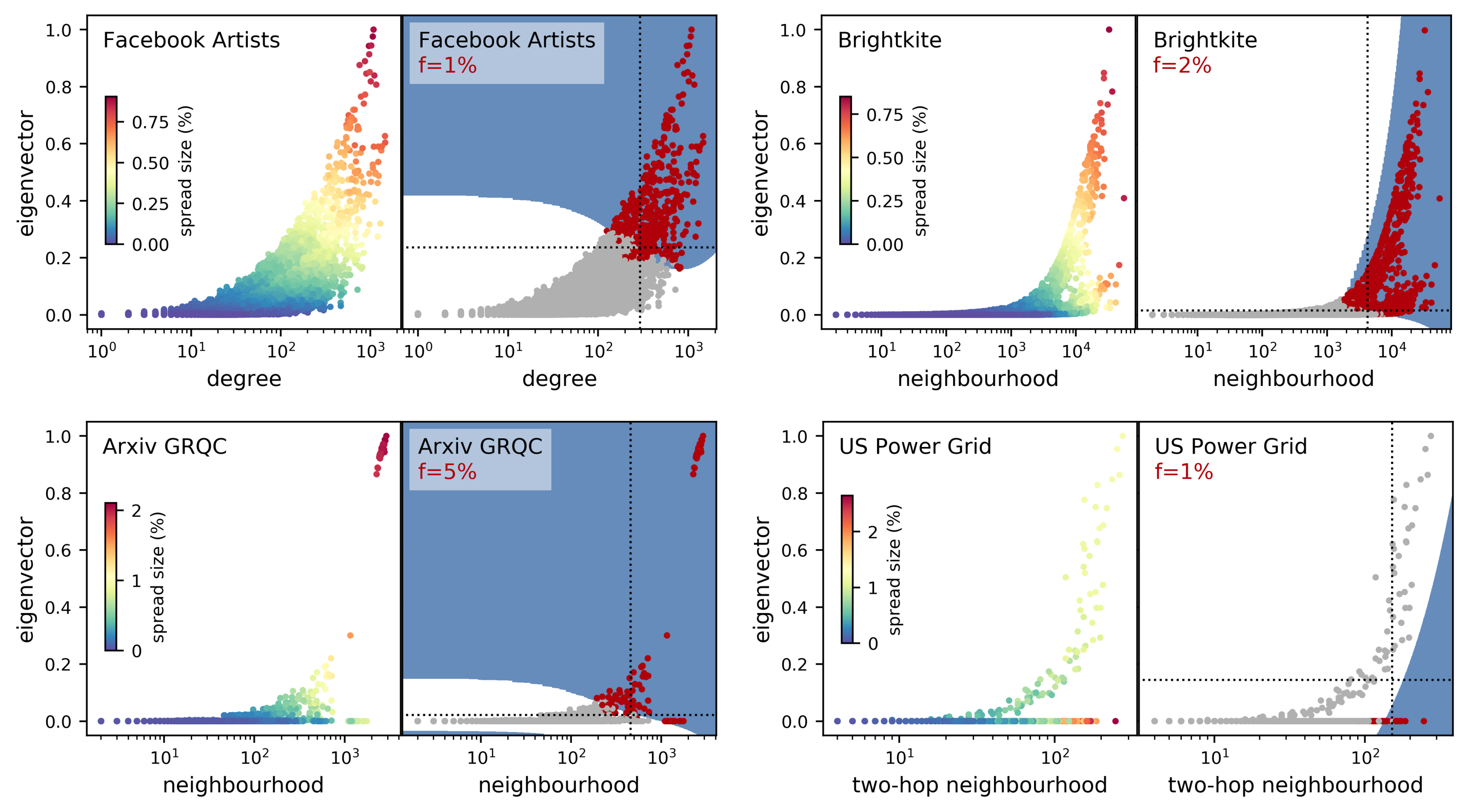}

    \caption{Network examples for which \textbf{eigenvector centrality combined with another centrality} improves the predictions of single-centrality rankers. In every left panel, a scatterplot of node centralities versus spread size. In every right panel, the top spreaders are coloured in red (or darker), the decision boundaries for rankers using either centrality are dotted lines, and the background colour shows the decision boundaries for the classifiers: a blue (or darker) background denotes the area predicted for top spreaders.}
    \label{fig:two-centrality-examples-Eigenvector}
\end{figure}

\paragraph{Combinations with the eigenvector centrality} We show four network examples in Fig.~\ref{fig:two-centrality-examples-Eigenvector}. For each network, the left panel maps the distribution of the spread size at the epidemic threshold for all the nodes in the network, against the pairing of the eigencentrality with a neighbourhood indicator. The right panel notes a value for $f$, and colours the nodes according to their true class: the red nodes are the top $f$ by spread size. Also in the right panel, two dotted lines show the decision boundaries made by the corresponding single-centrality rankers. If $f=1\%$, these boundaries are the $99$th percentiles for either centrality; a ranker will predict as top spreaders all nodes above this boundary. These ranking boundaries are improved upon by the classifier, whose decision boundary is shown as the transition between background colours, with a blue (or darker) background showing the centrality space where the top spreaders are predicted to be. (Note that only part of this centrality space may be occupied by nodes; in other words, not every combination of centrality values may be physically possible.) The optimal decision boundary would leave no nodes misclassified and would lead to values of 1 for both the precision function and the recall or F1 score.

There are clear commonalities among the improved decision boundaries in Fig.~\ref{fig:two-centrality-examples-Eigenvector}: for Facebook Artists, Brightkite, and Arxiv GRQC, the \emph{joint} increase in the values of both centralities in the pair is what determines an effective spreader. For Facebook Artists and Brightkite (both relatively large networks of over 50,000 nodes), ranking the nodes by only one centrality would place some nodes in the wrong class; unlike this, the two-centrality classifier (F1 scores of 0.920 and 0.924, respectively) draws a decision boundary that is much closer to optimal. We illustrated the intuition behind the Arxiv GRQC result (F1 score 0.900) in Fig.~\ref{fig:example-Arxiv_GRQC}: the size of the local neighbourhood does affect the spreading ability of nodes, but proximity to the `hub' of high eigencentrality also helps.

There are also exceptions from this. The US Power Grid network (4,941 nodes) shown in the same figure has an outlying cluster of low-eigencentrality nodes as top spreaders, while the lesser spreaders instead follow the expected trend described above. Supplementary Figure S1 shows the cause: a small hub of high eigencentrality values lies at a periphery of the network, while a larger region of nodes with large neighbourhoods (but low eigencentrality) is located far apart. It is the latter, larger region which enables the top 1\% of the spreaders, and the classifier is able to learn this pattern slightly better, with a 0.162 increase (F1 score 0.509) compared to the $r(f)$ of ranking by the two-hop neighbourhood alone.

\begin{figure}[htb]

    \includegraphics[width=\textwidth]{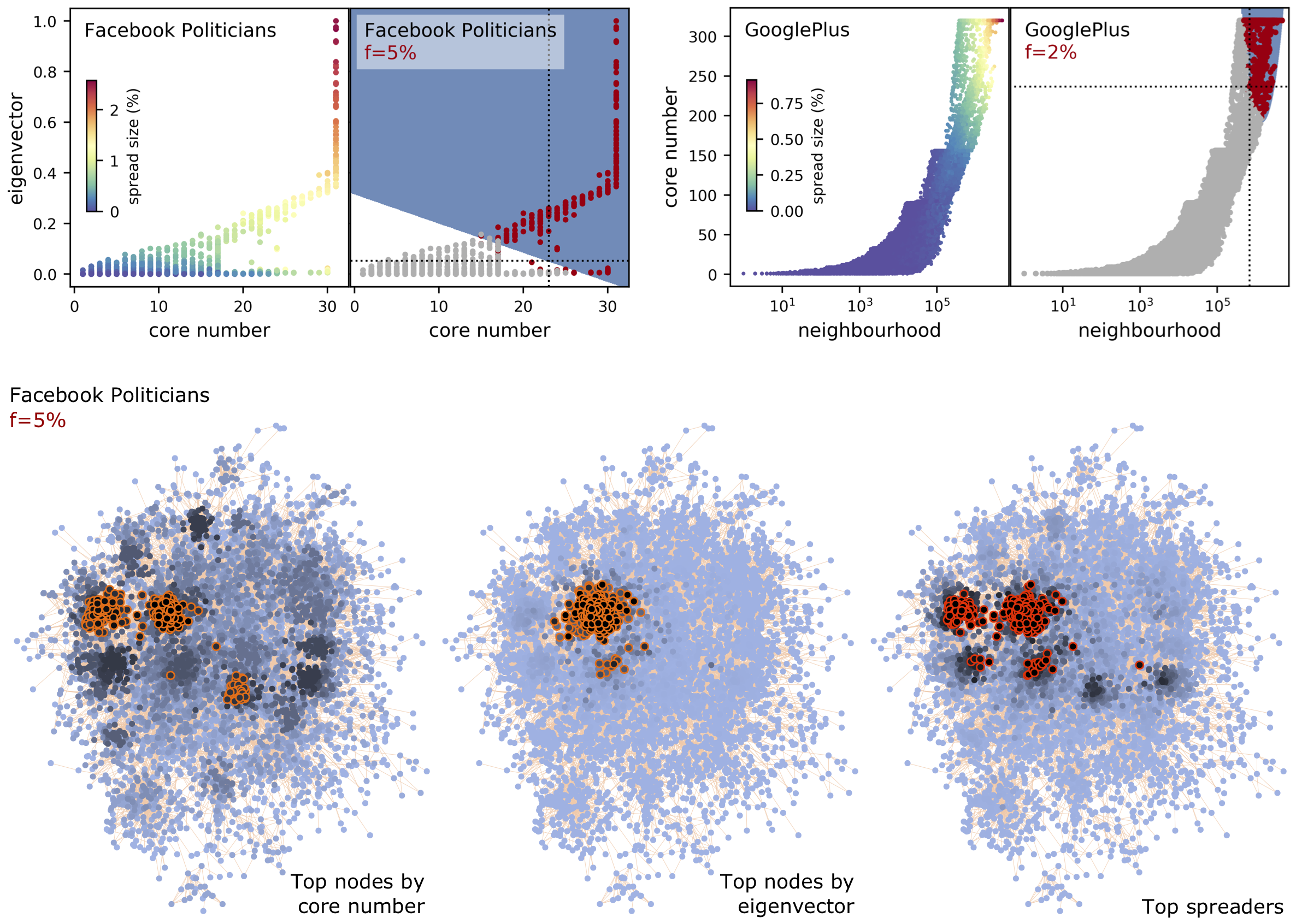}

    \caption{As Fig.~\ref{fig:two-centrality-examples-Eigenvector}, for \textbf{core number combined with another centrality}.}
    \label{fig:two-centrality-examples-CoreNumber}
\end{figure}

\paragraph{Combinations with the core number} A similar intuition holds when pairing the core number with eigenvector centrality, and also with neighbourhood centralities. (Other pairings with the core number are less effective.) We show two examples in Fig.~\ref{fig:two-centrality-examples-CoreNumber}. Again it is the joint increase in both centralities which enables superspreading. For Facebook Politicians (F1 score 0.894), Fig.~\ref{fig:two-centrality-examples-CoreNumber} (bottom) also illustrates the intuition. A number of dense cores are distributed in the network, with the highest core numbers not in close proximity, but isolated by regions of low density. On the other hand, a single region of high eigencentrality exists, and the top 5\% of spreaders are located exactly in those cores of highest eigencentrality. Interestingly, pairing the core number with a neighbourhood centrality (GooglePlus, F1 score 0.968) also shows that not all the nodes in dense cores are equally good spreaders, and that their neighbourhood size can help to make a selection.

\begin{figure}[htb]

    \includegraphics[width=\textwidth]{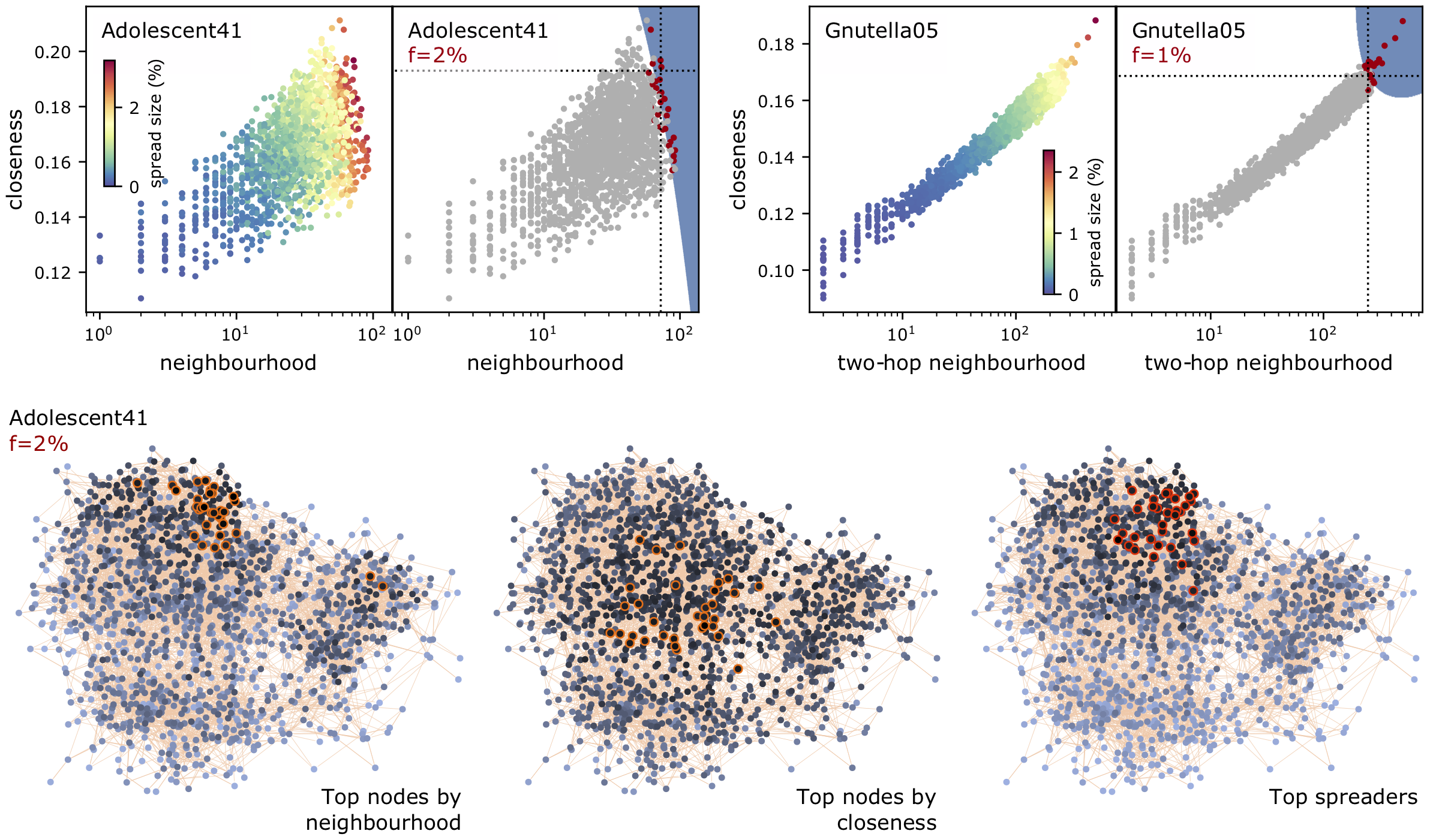}

    \caption{As Fig.~\ref{fig:two-centrality-examples-Eigenvector}, for \textbf{closeness combined with another centrality}.}
    \label{fig:two-centrality-examples-Closeness}
\end{figure}

\paragraph{Combinations with closeness} Closeness also plays a role similar to the eigencentrality---that of guiding the selection of nodes away from more peripheral nodes with dense neighbourhoods, towards the centre of the network, with an increase in performance. Figure~\ref{fig:two-centrality-examples-Closeness} shows two examples. In the Adolescent41 offline social network (1,640 nodes), the best ranker is that by neighbourhood ($r(f)=0.469$), but when considering also closeness, the F1 score rises to 0.598. On the topology of the network (at the bottom of the same figure), closeness values identify only very few of the top spreaders, while the neighbourhood size identifies more; the correct top spreaders, however, again lie in a region where both centralities jointly have high values. In the Gnutella05 computer network, for a similar reason, the best ranker is instead closeness ($r(f)=0.594$), but when considering also the two-hop neighbourhood, the F1 score rises to 0.725.

In the examples from Figs.~\ref{fig:two-centrality-examples-Eigenvector} to \ref{fig:two-centrality-examples-Closeness}, each classifier's decision boundary improves upon the decision boundary of the best ranker such that $r(f)$ is raised by between 0.090 and 0.213. Among our 60 test cases, we also found other examples of networks, combined with certain values for $f$, for which the single-centrality rankers could not be improved by any classifier. For example, only when $f=1\%$, none of the five Adolescent networks is resolved any better by using two centralities---but also there the performance improves when $f$ increases. 

\begin{figure}[htb]
    \centering

    \includegraphics[width=.8\textwidth]{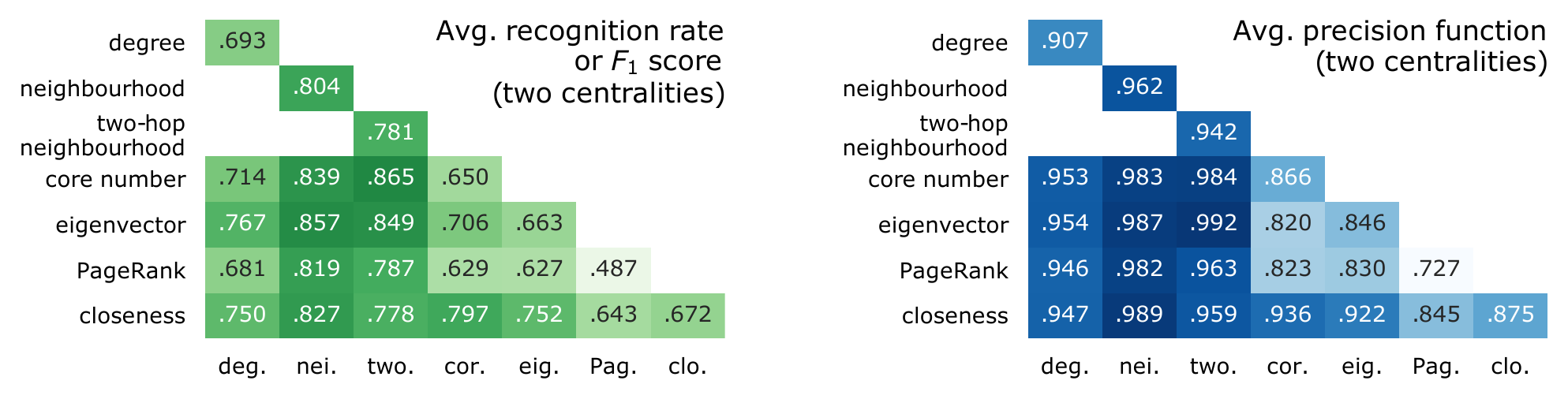}

    \caption{The success of \textbf{single and pairs of centralities} at predicting spreaders: for each pair of centralities, the average performance score across all networks and values of $f$. The diagonal is the result of ranking by a single centrality and it is scored by the recognition rate and the precision function. The rest of the matrix is the result of classification by two centralities and is scored by the F1 score and the precision function.}
    \label{fig:twocentrality-summary}
\end{figure}

From all pairs of centralities, the combination of two-hop neighbourhood and core number has the best average F1 score (0.865) across all the network cases in this study, and across the range of $f$. On the other hand, the combination of two-hop neighbourhood and eigenvector has the best average precision function (0.992). Figure~\ref{fig:twocentrality-summary} is a summary for the averages of both performance scores across all single centralities (on the diagonal) and pairs of centralities (the rest of the matrix). All possible pairs of centralities are studied, except for the redundant combinations between degree and neighbourhood, and between the two types of neighbourhood centralities. The six pairs which improve significantly on the most predictive ranker are all composed of one of the neighbourhood centralities, and one of: core number, eigenvector centrality, closeness, or PageRank. These six pairs improve on both recall and precision function.

% ______________________________________________________________________________________________________________
\subsection*{Multi-centrality predictors and summary of results}

While the previous subsection demonstrated that centrality indicators can play on each others' strengths and improve the prediction of top spreaders by the SIR diffusion model at the critical threshold, we now show that classifiers using all seven centralities as features give near-perfect prediction on most network examples. One exception is that of offline human social networks (the HS network category) and only at very low fractions $f$. This category contains networks that are not structurally unusual, but are some of the smallest networks in the study, which leads to very few training data points, thus lower classification performance.

\begin{figure}[htb]

    \includegraphics[width=\textwidth]{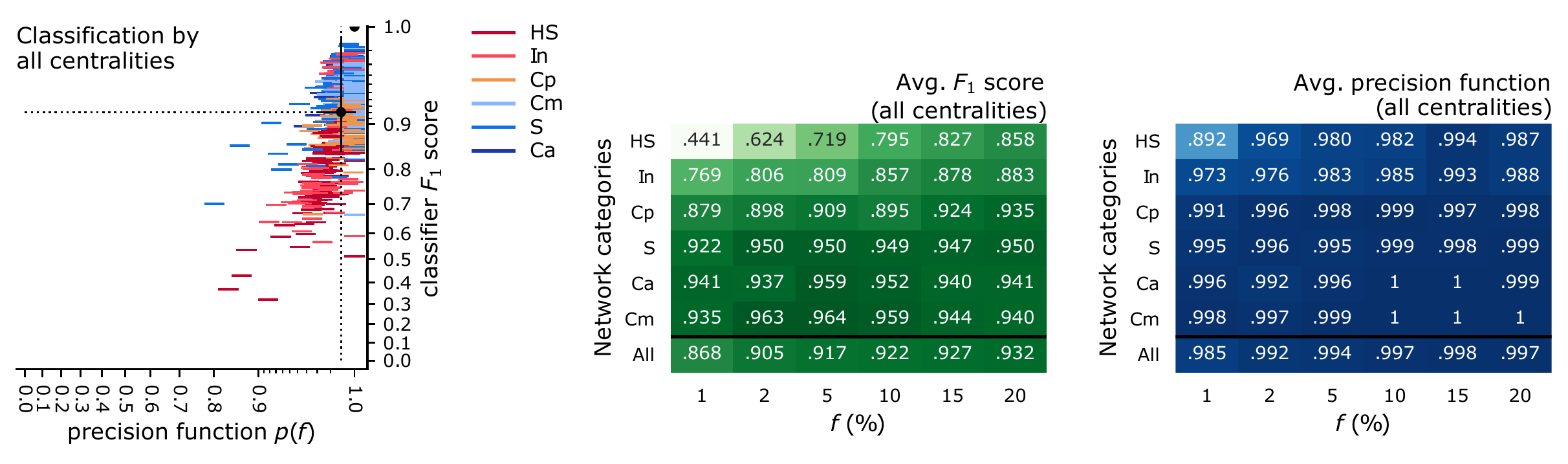}

    \caption{The success of \textbf{classifiers using all centralities} at predicting spreaders, across all networks and values of $f$, at the SIR epidemic threshold. (left) Each data point denotes a prediction task, and the colour shows the category of the network (listed in Table~\ref{tab:networks}). The centroid of the point cluster and the standard deviation on both axes are marked (counterpart to Fig.~\ref{fig:twometric-comparison-ranking}). (right) The average performance scores across all networks in one of six network categories, and across all values of $f$ (counterpart to Fig.~\ref{fig:twocentrality-summary}).}
    \label{fig:allcentralities-summary-SVM12-critical}
\end{figure}

We train a seven-centrality SVM classifier for each prediction task, and summarise the results in Fig.~\ref{fig:allcentralities-summary-SVM12-critical}. The centroid of all prediction scores (Fig.~\ref{fig:allcentralities-summary-SVM12-critical}, left) is an average recognition rate of 0.921, and an average precision function of 0.995. While the precision function was almost as high (0.992) when training the classifier using only the eigenvector centrality and the two-hop neighbourhood as features (Fig.~\ref{fig:twocentrality-summary}), the average recognition rate is now further improved by adding more features to the statistical model. Not all six network categories are equal: a breakdown of the scores by network category and by the value of the fraction $f$ (Fig.~\ref{fig:allcentralities-summary-SVM12-critical}, right) shows that recognising the top 1\% of spreaders in the Adolescent networks (the HS network category) remains difficult. All other prediction tasks are resolved well, particularly when performance is measured by the precision function, which ranges between 0.969 and 1.

These conclusion hold also above the epidemic threshold, at $1.5\cdot\lambda_c$; numerical results showing very similar prediction scores are in Supplementary Fig. S2. They are also not an artefact of the type of statistical model used in the classifier. When training nonlinear Random Forest classifiers, which are high-variance so---in general---are able to obtain better performance than the polynomial SVM, a similar conclusion emerges (Supplementary Fig. S3), so there no significant advantage to using higher-variance classifiers.

% ______________________________________________________________________________________________________________
\section*{Discussion}

\paragraph{Insights gained} We showed that two or more classical centrality indicators contain sufficient statistical information about the nodes in a real-world network to train an accurate predictor of SIR influence, and outperform node rankers. The decision boundaries between the two classes, as learnt by classifiers, demonstrate where the advantage of multi-variate prediction comes from: certain centrality indicators are particularly good complements to others. Notably, there are multiple answers to the question: what is a good pair of centralities? For the degree centrality, the best complement is the eigenvector centrality. For the neighbourhood centrality (the best overall single ranker), three other centralities make good complements: the eigenvector centrality, closeness, and core number (with PageRank also close). For those network cases where multi-variate prediction has an advantage, the joint distribution of the centralities and the SIR influence is such that one centrality (or, a one-dimensional decision boundary) is insufficient to classify the nodes accurately, but a multi-dimensional decision boundary is able to refine the decision in the most important region of centrality values. When seven classical centralities are used, the prediction performance is close to optimal (to an average recognition rate of 0.921, and an average precision function of 0.995).

We showed the topological intuition behind this improvement in the prediction of superspreaders. Often, when a subset of the top nodes by local centrality indicators are located in more peripheral regions of the network, global centrality indicators step in and act as a selector and guide towards the effective centre of the network, so that the nodes selected jointly maximise the values of both centralities. In exceptional topologies, when the global centrality has high values at a peripheral location (such as US Power Grid, in Supplementary Fig. S1), the roles reverse: the local centrality becomes the selector, and the statistical model learns that high global centrality values are not beneficial.

\paragraph{Practical use, assumptions, and limitations} The basic insight of jointly maximising the values of two or more centralities can help improve existing, unsupervised node ranking methods. The advantage of ranking algorithms is that they are unsupervised, i.e., require no ground truth; their disadvantage is lower recall and precision. 

Network practitioners can also use supervised classification as presented here, and train a new classifier on a new network. While this method delivers good predictions, it assumes (a) complete knowledge of the network links, and (b) means to estimate the spread size for a fraction of the network nodes. If historical diffusion data is available (such as the number of retweets on Twitter), this data replaces the need to simulate a theoretical diffusion model in order to obtain ground truth for the spread size. Only a fraction of nodes need ground truth data, since the statistical classifier is trained on a random sample of the nodes in the network, and will predict the class for the others. The size of the training data necessary to obtain good predictions depends on the network and on the distributions of centrality and influence values, but is expected to be small. In Supplementary Fig. S4, we measure the required training set size via plotting learning curves for three of the largest networks in this study. These show that, to obtain maximum performance, some networks only require a training data size of 1\% of the network size, while others need around 10\%. The set of centralities to use as features can be tailored to the computational budget available. The type of statistical model can also be tailored with the network size: heuristic training algorithms, such as those training Random Forest classifiers, scale better with large networks.

\paragraph{Future work} There are follow-ups to explore as continuations of this study, at the intersection between real-world network dynamics and machine learning. A method to train a single statistical model for predicting superspreaders across networks is desirable, as long as its performance remains good; this was previously achieved only for small networks\cite{bucur2020beyond}. Other directions include the prediction of other measures of node influence (such as the measured diffusion of information in large online social networks\cite{pei2014searching}) and of node importance (such as the ability of a node to block the diffusion of information), and the study of other types of networks (such as different network categories, networks with node attributes, and networks with dynamic structure).

% ______________________________________________________________________________________________________________
\section*{Methods}

% ______________________________________________________________________________________________________________
\subsection*{Networks, centrality indicators, and the estimation of node influence}

Most of the networks used as case studies (see Table~\ref{tab:networks} for the overview) model entire communities at a specific point in time. This is the case for the high-school friendships in the Adolescent networks, the daily Gnutella peer-to-peer file sharing networks, the five sets of institutional email exchanges, or the networks of mutual likes between verified Facebook pages. A minority of the networks (such as the Facebook Stanford friendships, collected from survey participants) are instead bounded samples from a larger community. All are (transformed into) directed, strongly connected, and unweighted networks; when the original version in the repository had timestamp, attribute, or weight annotations, these were removed. The direction of the edges is reversed when needed, to model information flow---so the degree centrality of interest is the out-degree. To be able to study the closeness centrality\cite{newman2018networks} which computes the lengths of shortest paths, only the largest strongly connected component (SCC) was kept. These networks were selected from public repositories such that (a) they fit into these six categories, and (b) have the size of their SCC above 1,000 nodes. The upper bound on network size is simply imposed by finite computing resources.

\begin{table}[htb]
\footnotesize
\centering
\begin{tabular}{lllrl}
\textbf{Cat.} & \textbf{Rep.} & \textbf{Network} &  \textbf{Size}  &  $\lambda_c$  \\ 
\hline
\textit{HS} & \textit{K} & Adolescent36 & 1,671 & 0.29 \\
\textit{HS} & \textit{K} & Adolescent40 & 1,679 & 0.23 \\
\textit{HS} & \textit{K} & Adolescent41 & 1,640 & 0.26 \\
\textit{HS} & \textit{K} & Adolescent49 & 1,149 & 0.25 \\
\textit{HS} & \textit{K} & Adolescent50 & 2,155 & 0.25 \\
\textit{S}  & \textit{K} & Advogato & 3,140 & 0.041 \\
\textit{Ca} & \textit{S} & Arxiv Astro & 17,903 & 0.0155 \\
\textit{Ca} & \textit{S} & Arxiv CondMat & 21,363 & 0.045 \\
\textit{Ca} & \textit{S} & Arxiv GRQC & 4,158 & 0.080 \\
\textit{Ca} & \textit{S} & Arxiv HEPPh & 11,204 & 0.008 \\
\textit{Ca} & \textit{S} & Arxiv HEPTh & 8,638 & 0.0925 \\
\textit{Cp} & \textit{S} & AS CAIDA 20040105 & 16,301 & 0.033 \\
\textit{Cp} & \textit{S} & AS CAIDA 20041206 & 18,501 & 0.028 \\
\textit{Cp} & \textit{S} & AS CAIDA 20051205 & 20,889 & 0.028 \\
\textit{Cp} & \textit{S} & AS CAIDA 20061225 & 23,918 & 0.027 \\
\textit{Cp} & \textit{S} & AS CAIDA 20071112 & 26,389 & 0.030 \\
\textit{S}  & \textit{S} & Brightkite & 56,739 & 0.0185 \\
\textit{Cm} & \textit{S} & Email Enron & 33,696 & 0.012 \\
\textit{Cm} & \textit{S} & Email EU & 34,203 & 0.022 \\
\textit{Cm} & \textit{K} & Email Linux & 18,531 & 0.0075 \\
\textit{Cm} & \textit{M} & Email UCL & 12,625 & 0.035 \\
\textit{Cm} & \textit{K} & Email URV & 1,133 & 0.070 \\
\textit{S}  & \textit{S} & Epinions & 32,223 & 0.0135 \\
\textit{In} & \textit{K} & Euroroad & 1,039 & 1.3 \\
\textit{S}  & \textit{S} & Facebook Artists & 50,515 & 0.007 \\
\textit{S}  & \textit{S} & Facebook Athletes & 13,866 & 0.030 \\
\textit{S}  & \textit{S} & Facebook Companies & 14,113 & 0.057 \\
\textit{S}  & \textit{S} & Facebook Government & 7,057 & 0.014 \\
\textit{S}  & \textit{M} & Facebook New Orleans & 63,392 & 0.0098 \\
\textit{S}  & \textit{S} & Facebook Politicians & 5,908 & 0.031 \\
\hline
\end{tabular}
\hspace{15mm}
\begin{tabular}{lllrl}
\textbf{Cat.} & \textbf{Rep.} & \textbf{Network} &  \textbf{Size}  &  $\lambda_c$  \\ 
\hline
\textit{S}  & \textit{S} & Facebook Public Figures & 11,565 & 0.020 \\
\textit{S}  & \textit{S} & Facebook Stanford & 4,039 & 0.011 \\
\textit{S}  & \textit{S} & Facebook TV Shows & 3,892 & 0.049 \\
\textit{S}  & \textit{S} & GitHub & 37,700 & 0.0105 \\
\textit{Cp} & \textit{S} & Gnutella04 & 4,317 & 0.29 \\
\textit{Cp} & \textit{S} & Gnutella05 & 3,234 & 0.32 \\
\textit{Cp} & \textit{S} & Gnutella24 & 6,352 & 0.39 \\
\textit{Cp} & \textit{S} & Gnutella25 & 5,153 & 0.42 \\
\textit{Cp} & \textit{S} & Gnutella30 & 8,490 & 0.35 \\
\textit{Cp} & \textit{S} & Gnutella31 & 14,149 & 0.38 \\
\textit{S}  & \textit{S} & GooglePlus & 69,501 & 0.0019 \\
\textit{S}  & \textit{K} & Hamsterster & 2,000 & 0.029 \\
\textit{Ca} & \textit{M} & IMDB & 47,719 & 0.003 \\
\textit{In} & \textit{K} & OpenFlights & 3,354 & 0.024 \\
\textit{S}  & \textit{K} & PGP & 10,680 & 0.065 \\
\textit{S}  & \textit{S} & Twitch DE & 9,498 & 0.0085 \\
\textit{S}  & \textit{S} & Twitch EN & 7,126 & 0.033 \\
\textit{S}  & \textit{S} & Twitch ES & 4,648 & 0.014 \\
\textit{S}  & \textit{S} & Twitch FR & 6,549 & 0.0098 \\
\textit{S}  & \textit{S} & Twitch RU & 4,385 & 0.0185 \\
\textit{S}  & \textit{S} & Twitch PT & 1,912 & 0.013 \\
\textit{S}  & \textit{S} & Twitter Stanford & 68,413 & 0.0115 \\
\textit{In} & \textit{K} & US Airports & 1,402 & 0.020 \\ 
\textit{In} & \textit{K} & US Power Grid & 4,941 & 0.87 \\
\textit{Cm} & \textit{K} & WikiTalk AR & 8,797 & 0.018 \\
\textit{Cm} & \textit{K} & WikiTalk IT & 36,356 & 0.008 \\
\textit{Cm} & \textit{K} & WikiTalk NL & 18,598 & 0.012 \\
\textit{Cm} & \textit{K} & WikiTalk PT & 21,747 & 0.009 \\
\textit{Cm} & \textit{K} & WikiTalk RU & 22,664 & 0.011 \\
\textit{Cm} & \textit{K} & WikiTalk ZH & 10,831 & 0.013 \\
\hline
\end{tabular}
\caption{\label{tab:networks} 
The 60 case studies. We use the largest strongly connected component, whose size (node count) is reported here. \textbf{Rep.} denotes the source repository: \textit{K} (KONECT\cite{konectweb,kunegis2013konect}), \textit{M} (H. Makse\cite{makseweb}), or \textit{S} (SNAP\cite{snapnets}). \textbf{Cat.} is the KONECT category most suited to the case study: 
\textit{Ca} (Coauthorship),
\textit{Cm} (Communication between people),
\textit{Cp} (Computer),
\textit{HS} (Human Social, offline),
\textit{In} (Infrastructure),
\textit{S} (Social, online).
$\lambda_c$ denotes the epidemic threshold, estimated numerically.
} % Legend: 350 words max
\end{table}

The following centrality indicators were computed for every node in every network: its \emph{degree}, \emph{neighbourhood} (i.e., the sum of the degrees of the nearest neighbours, previously denoted $k_{\mathit{sum}}$ and found to be a competitive predictor in a previous study\cite{pei2014searching}), \emph{two-hop neighbourhood} (as before\cite{pei2014searching} for nearest neighbours exactly two hops away and previously denoted $k_{\mathit{2sum}}$), \emph{PageRank}\cite{newman2018networks} with a 0.85 damping factor, \emph{eigenvector centrality}\cite{newman2018networks}, \emph{closeness centrality}\cite{newman2018networks}, and \emph{core number}\cite{kitsak2010identification}. An additional set of indicators that we tried, the \emph{link strength} of a node towards upper, equal, or lower shells\cite{liu2015core}, denoted $r^u, r^e$, or $r^l$, did not provide notable results.

The ultimate influence of a node in a network is estimated numerically, as the average among $10^4$ runs of the susceptible– infectious–recovered (SIR)\cite{anderson1979population} diffusion model for infectious diseases. In SIR, an infectious node infects a susceptible neighbour at a \emph{rate} $\beta$ (meaning the number of infection events per time unit, so can be higher than 1). An infectious node recovers at a rate $\mu$. The effective transmission rate is $\lambda=\beta/\mu$. Here, we take $\mu=1$ and study the normalized rate $\lambda$. 

As $\lambda$ increases in SIR simulations, the size of the outbreaks increase from an infinitesimal fraction to a finite fraction of the network size. The regime of interest is neither very low $\lambda$ values (in which case, the diffusion remains localised to the neighbourhood of the seed node) nor very high (in which case, all nodes should reach a large fraction of the network). Since our test cases are both finite in size, and diverse (a scenario studied previously\cite{shu2015numerical}), we estimate the \emph{epidemic threshold} $\lambda_c$ numerically by identifying it with the variability measure\cite{shu2015numerical} $\Delta = \frac{\sqrt{\langle\rho^2\rangle - \langle\rho\rangle^2}}{\langle\rho\rangle}$. Here, $\rho$ denotes the random variable of outbreak size from different seed nodes, and $\langle\cdot\rangle$ denotes the mean. Given a value for $\lambda$, $\Delta$ is estimated by setting seed nodes from a random sample of $10^4$ of the nodes in a network (or the entire network size, if this is smaller). After estimating $\Delta$ for a range of $\lambda$ values at regularly spaced intervals, we take $\lambda_c$ to be the position of the peak of $\Delta$. The resulting values are noted in Table~\ref{tab:networks}. The maximum spread size (influence) at $\lambda_c$ in any network is between 0.7\% and 6\% of the network size (with two exceptions among the smallest infrastructure networks, where this reaches 8\% and 11\%). 

% ______________________________________________________________________________________________________________
\subsection*{Ranking by a single centrality}

\paragraph{Method} 
We first predict superspreaders using the single-centrality ranking method common in prior studies\cite{kitsak2010identification,pei2014searching,de2014role,liu2015core,macdonald2012spreaders}, and also carry forward the performance metrics defined in these studies. This ranking method builds the assumption that higher centrality values for a node will also indicate higher node influence. Given a centrality $C$, first all the nodes have their values for $C$ computed. The top fraction $f$ of spreaders is then predicted to be the fraction $f$ of nodes with the highest values for $C$. At ties between nodes (which occur for discrete-valued centralities such as degree and core number) a random subset of the tied nodes are selected. This random sampling is then repeated $10^2$ times for a bootstrap technique (described below), which averages among the scores of these individual random choices. 

\paragraph{Performance metrics} 
In prior studies, this ranking is evaluated via two metrics. Denote by $I_f$ the set of the top fraction $f$ of nodes as ranked by their SIR influence, and by $C_f$ the set of top fraction $f$ of nodes as ranked by their centrality values; the sizes of these sets are equal for a given $f$, $\left| I_f\right| = \left| C_f\right|$. Also denote by $\rho_i$ the spread size when setting node $i$ as seed. The \textit{recognition rate} $r(f)$ measures the extent to which the identities of the predicted superspreaders match the true identities\cite{pei2014searching}. A synonym for the recognition rate is \emph{recall}. The \textit{precision function} $p(f)$ is a weaker, but more practically useful performance measure comparing the spread of the predicted superspreaders to that of the true top spreaders:
\begin{equation}
    r(f) = \frac{\left| I_f \cap C_f \right|}{\left| I_f \right|} 
    \qquad \text{and} \qquad
    p(f) = \frac{\text{avg}_{i\in C_f} \rho_i}{\text{avg}_{i\in I_f} \rho_i}
    \label{eq:pm1}
\end{equation}
Both metrics take values in the interval $[0, 1]$. An imprecision function $\epsilon(f)$ was defined previously\cite{kitsak2010identification}, such that lower values of $\epsilon(f)$ are better. Here, to present the two metrics in a unified fashion, we use instead $p(f) = 1-\epsilon(f)$, such that higher values are better for both $r(f)$ and $p(f)$. A confidence interval was originally provided for $r(f)$ by bootstrap\cite{pei2014searching}. Here, we apply a bootstrap technique when estimating both metrics. Given a network of $N$ nodes, $10^2$ times, we draw a random sample of the $N$ nodes uniformly with replacement. Among these nodes, the ranking method is applied and a prediction is made and evaluated via either $r(f)$ or $p(f)$, as needed. The final value for each performance metric is the average, together with the 95\% confidence interval among these samples.

% ______________________________________________________________________________________________________________
\subsection*{Classification by a combination of centralities}

\paragraph{Method} % SVM, training, testing, interpretation
A multi-centrality method learns a \emph{discriminative statistical model} able to classify network nodes into superspreaders or not. For this, a dataset is formed for every network; a record describes a node via its centrality values (the predictors). When training the model to recognise the top fraction $f$ of the nodes, the nodes are ranked by their true SIR spread size, and each node is assigned one of two target classes based on whether or not they are in the top fraction $f$. The model is trained and tuned on a training fraction $t=0.5$ of the nodes (sampled randomly without replacement), and tested on the remaining nodes.

A binary statistical classifier learns a decision boundary between the classes. We use a \emph{support-vector machine} (SVM)\cite{ben2001support}, which learns optimal separating hyperplanes in the multi-dimensional predictor space, including in cases where the classes overlap in this space. Here, the optimal decision boundary is that which leaves the largest margin in space between the classes, with still allowing some data points to fall on the wrong side of the boundary. SVMs have advantages: (a) they are optimal learners rather than heuristics, and (b) the kernel function $K$ and the regularisation parameter $C$, which ultimately give the shape and variance of the boundary\cite{hastie2009elements}, are tunable hyperparameters. 

We aim to obtain the simplest, most interpretable classifier with good performance; higher-variance classifiers bring little performance advantages for this problem, and may lose in interpretability. The results presented are for second-degree polynomials $K$ (which gives a low-variance model, less prone to overfitting), $C$ tuned in the range $[1, 100]$ with five-fold cross-validation, and a fixed tolerance for the stopping criterion\cite{scikit-learn} of 5e-4. No class weights are added to balance the classes artificially. (We tested other, higher-variance statistical models: SVMs with third-degree polynomials for $K$, and models based on decision trees, either boosted or in ensembles\cite{breiman2001statistical}; since they had similar performance to the SVM with a second-degree polynomial for kernel, we retain and present the results for the latter.) We show the decision boundaries learnt by two-centrality models via plotting them in the predictor space.

\paragraph{Performance metrics}
For a network of size $N$ and the fraction $f$, a classifier produces a guess for the class of each network node in the test set. We port the same notation $C_f$ to mean here the set of nodes classified as top spreaders. The number of superspreaders predicted in this way is decided by the classifier, and may not equal $fN$. We measure the overlap between the classifier prediction and the ground truth with metrics similar to Eq.~\ref{eq:pm1}. In binary classification, the measure $r(f)$ as defined in Eq.~\ref{eq:pm1} is called \emph{recall} or sensitivity. It is a useful metric, but insufficient to characterise the classifier: alongside making many correct choices (giving a high true positive rate, $\left| I_f \cap C_f \right|$), the classifier may also add many false positives. The \emph{precision} metric helps to quantify the false positives, and a classical metric is the combination of recall and precision is their harmonic mean, the \emph{F1 score}\cite{van1979information}:
\begin{equation}
    \text{recall}(f) = r(f) = \frac{\left| I_f \cap C_f \right|}{\left| I_f \right|} 
    \qquad
    \text{precision}(f) = \frac{\left| I_f \cap C_f \right|}{\left| C_f \right|} 
    \qquad
    \text{F1 score}(f) = \frac{2}{\text{recall}(f)^{-1} + \text{precision}(f)^{-1}}
    \label{eq:pm2}
\end{equation}
Note that \emph{precision} is an established name in the area of Information Retrieval\cite{van1979information}, while the \emph{imprecision function} $\epsilon(f)$ which gave the \textit{precision function} $p(f)$ was defined recently\cite{kitsak2010identification} for analysing networks. Although the names are unfortunately too similar, their meaning is different and should not be confused.

The F1 score takes values in the interval $[0, 1]$. We apply to the classifier the second metric, the precision function $p(f)$, exactly as it is defined is Eq.~\ref{eq:pm1}. Its values can exceed $1.0$, in cases when the classifier predicts fewer than $fN$ superspreaders, and they are on average better than the true $fN$ superspreaders; we cap higher values to $1.0$. We estimate both F1 score and $p(f)$ by randomly drawing different training sets for the classifier (the same training fractions $t$ of the nodes) $10^2$ times, then training and testing the classifier on each draw. The final value for each performance metric is the average of the individual scores. % stdev not shown anywhere, so not mentioned

\bibliography{ms}

\section*{Author contributions statement}

% D.B. conceived and conducted the experiment(s), analysed the results, and wrote the manuscript. 
This contribution is single-authored.

\section*{Additional information}

\paragraph{Competing interests} The author declares no competing interests.

\end{document}

\flushbottom
\maketitle

\thispagestyle{empty}

%%%%%%%%%%%%%%%%%%%%%%%%%%%%%%%%%%%%%%%%%%%%%%%%%%%%%%%%%%%%%%%%%%%%%%%%%%%%%%%%%%%%%%%%%%%%%%%%%%%%%%%%%%%%%%%%

{   \centering
    \includegraphics[width=\textwidth]{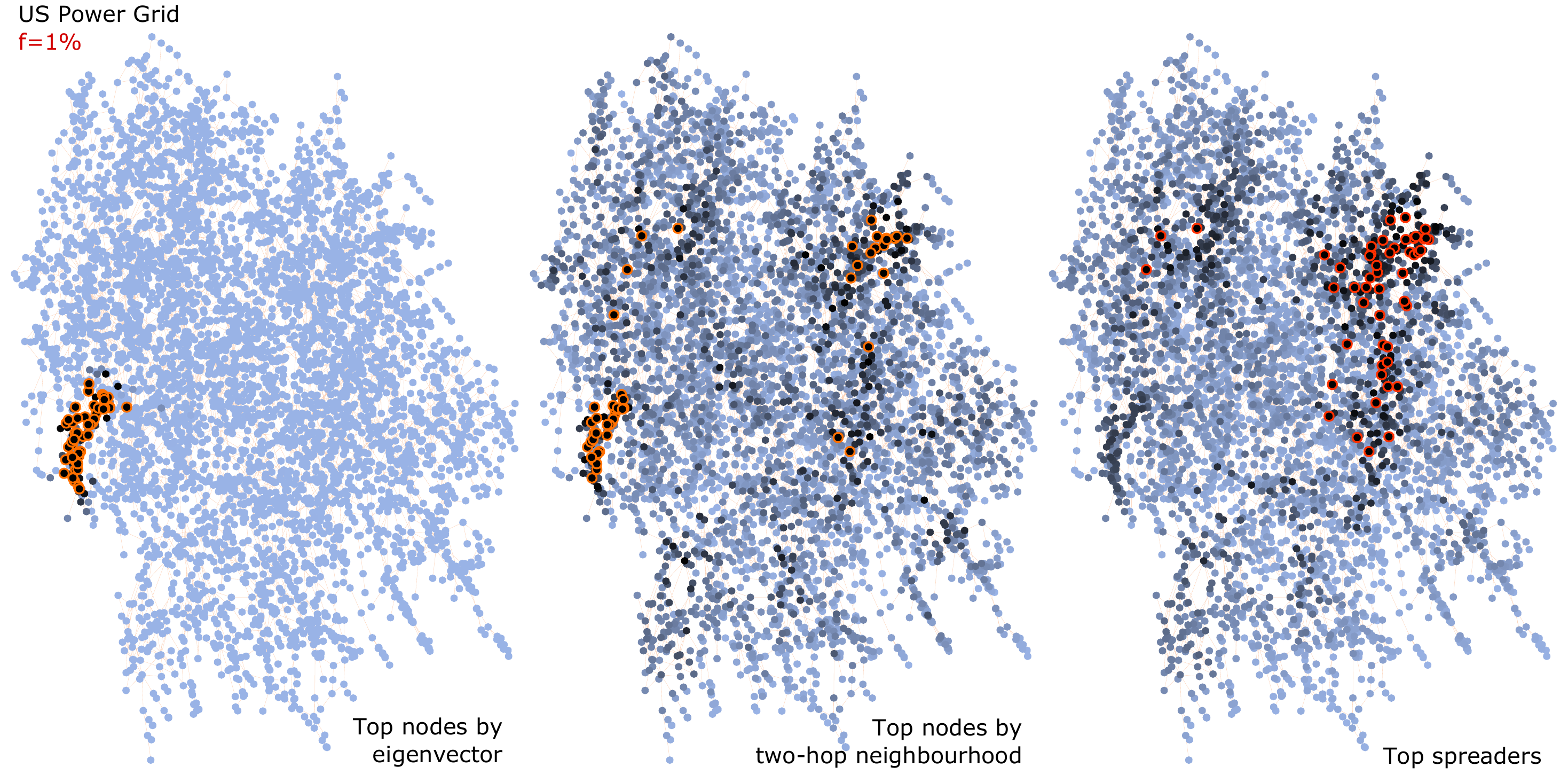}
}

\vspace{5mm}
\noindent\textbf{Supplementary Figure S1.} Comparing the location of the top nodes as ranked by (left) eigenvector centrality, (centre) two-hop neighbourhood size, and (right) SIR spread size at the epidemic threshold for the infrastructure network US Power Grid. The network layout is force-directed. The colour of the nodes in each panel shows the value of that metric: darker nodes have higher centrality values or spread size. The top $f=1\%$ of the nodes in each case are encircled.

\vspace{1cm}

{   \centering
    \includegraphics[width=\textwidth]{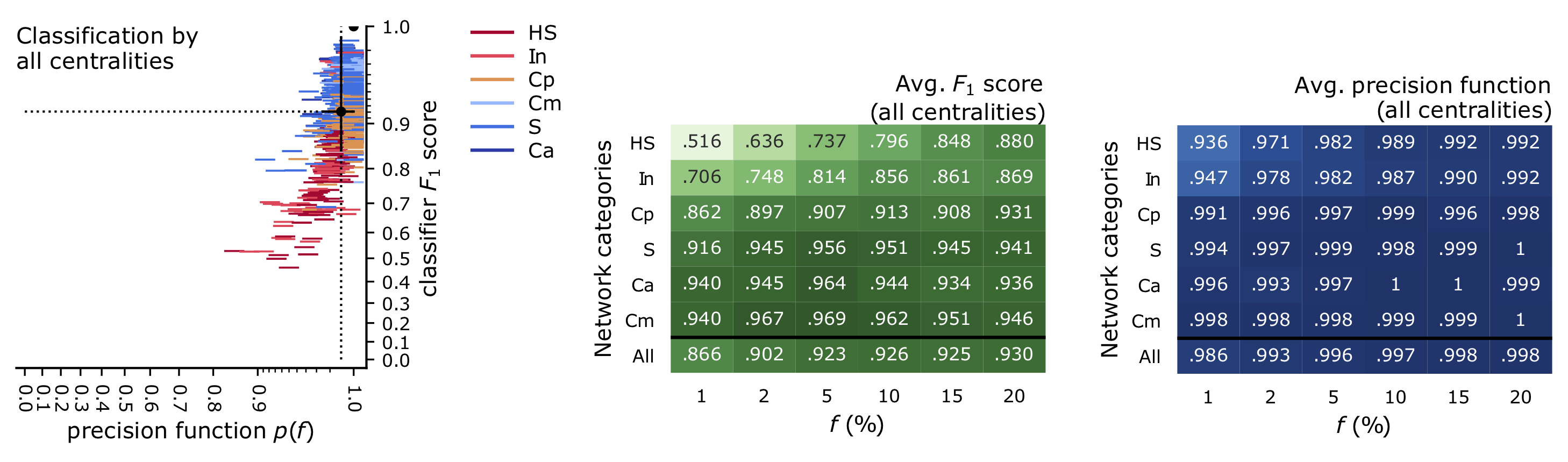}
}

\vspace{5mm}
\noindent\textbf{Supplementary Figure S2.} The success of \textbf{SVM classifiers using all centralities} at predicting spreaders, across all networks and values of $f$, \textbf{above the SIR epidemic threshold}, at $1.5\cdot\lambda_c$ for each network. (Counterpart to the summary results at the epidemic threshold $\lambda_c$ from Figure 7).

\vspace{1cm}

{   \centering
    \includegraphics[width=\textwidth]{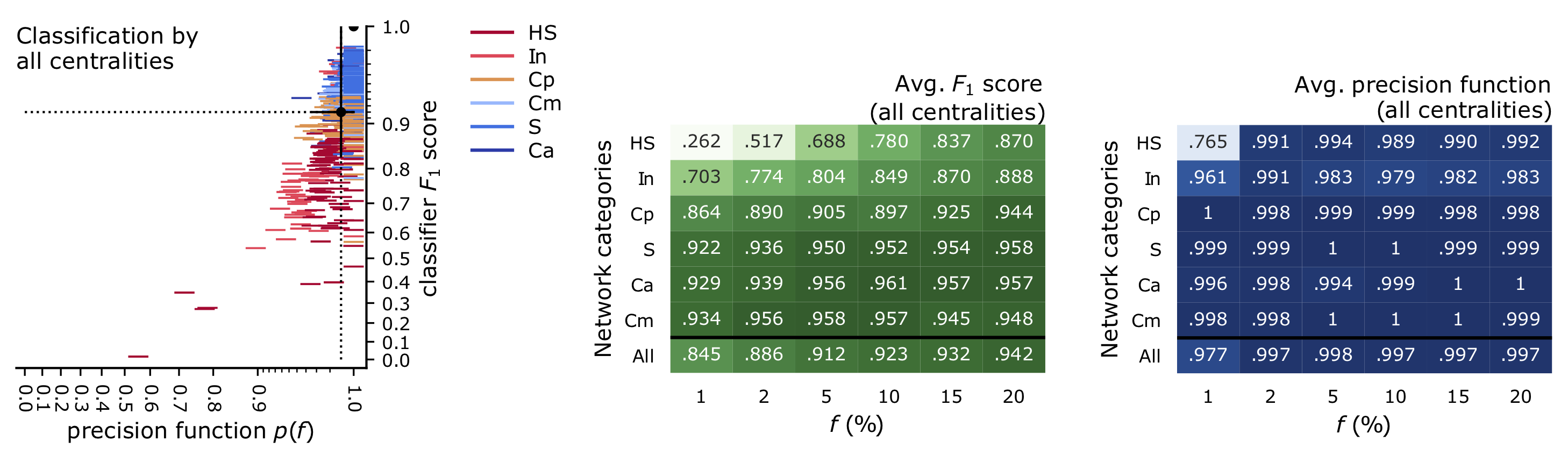}
}

\vspace{5mm}
\noindent\textbf{Supplementary Figure S3.} The success of \textbf{Random Forest classifiers using all centralities} at predicting spreaders, across all networks and values of $f$, \textbf{at the SIR epidemic threshold} $\lambda_c$ for each network. (Counterpart to the summary results for SVM classifiers at the epidemic threshold $\lambda_c$ from Figure 7).

\vspace{1cm}

{   \centering
    \includegraphics[width=.8\textwidth]{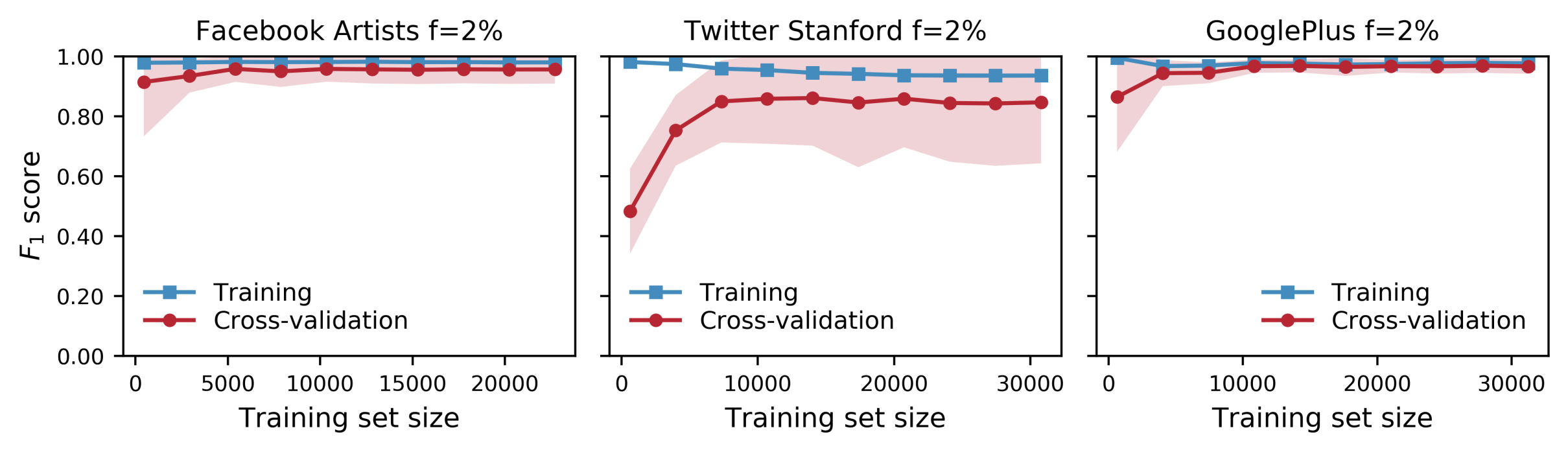}
}

\vspace{5mm}
\noindent\textbf{Supplementary Figure S4.} \textbf{Learning curves} for three large networks, showing the amount of training data (in terms of the number of nodes, on the x axis) versus the performance after training and cross-validation with all centralities (on the y axis). The training set size ranges between 1\% and 50\% of the network size, for each network. One can choose the size of the training data; a size is sufficient if the learniing curve shows a high cross-validation score, close to the training score.  

%%%%%%%%%%%%%%%%%%%%%%%%%%%%%%%%%%%%%%%%%%%%%%%%%%%%%%%%%%%%%%%%%%%%%%%%%%%%%%%%%%%%%%%%%%%%%%%%%%%%%%%%%%%%%%%%